\documentclass[aip,jcp,reprint,numerical,amsmath,amssymb,amsfonts,showkeys,showpacs]{revtex4-1}
\usepackage{bm,graphicx,xcolor,hyperref,rotating,lineno}
\newcommand{\mc}{\multicolumn}
\newcommand{\Ec}{E_{\rm c}}
\DeclareMathOperator{\Si}{Si}

\begin{document}

\title{Correlation energy of two electrons in a ball}

\author{Pierre-Fran\c{c}ois Loos}
\email{loos@rsc.anu.edu.au}
\author{Peter M. W. Gill}
\email{peter.gill@anu.edu.au}
\affiliation{Research School of Chemistry, Australian National University, Canberra, ACT 0200, Australia}
\date{\today}

\begin{abstract}
We study the ground-state correlation energy $\Ec$ of two electrons of opposite spin confined within a $D$-dimensional ball ($D \ge 2$) of radius $R$.  In the high-density regime, we report accurate results for the exact and restricted Hartree-Fock energy, using a Hylleraas-type expansion for the former and a simple polynomial basis set for the latter.  By investigating the exact limiting correlation energy $\Ec^{(0)} = \lim_{R \to 0} \Ec$ for various values of $D$, we test our recent conjecture [J. Chem. Phys. {\bf 131} (2009) 241101] that, in the large-$D$ limit, $\Ec^{(0)} \sim -\delta^2/8$ for any spherically-symmetric confining external potential, where $\delta=1/(D-1)$.
\end{abstract}


\maketitle

\section{\label{sec:intro} Introduction}

In the early days of quantum chemistry, there was considerable interest in cavity-confined atoms as a model for high-density atomic gas \cite{Michels37, Sommerfeld38, deGroot46}  
and extrapolation of high-density results provides a convenient but powerful route to understanding the intermediate-density regime. \cite{GellMann57}
Thanks to Hylleraas' work \cite{Hylleraas29}, the compressed helium-like ions have been widely studied \cite{Seldam52a,Gimarc67} and interest in these continues unabated. \cite{Aquino03, Aquino06, Aquino09, FloresRiveros10, Montgomery10}  Other confined systems such as electrons in square \cite{Alavi00,Ghosh05}, cylindrical \cite{Ryabinkin10} and spherical\cite{Thompson02,Jung03,Jung04,Thompson04a,Thompson04b,Thompson05} boxes have also attracted attention.  The last of these has been extensively used for the assessment of density-functional approximations \cite{Thompson02,Jung03,Jung04} and the study of Wigner molecules \cite{Wigner34} at low densities. \cite{Thompson04a,Thompson04b,Thompson05}

In a previous article, \cite{EcLimit09} we studied the high-density correlation energy $\Ec^{(0)}$ for various two-electron systems confined to a 
$D$-dimensional space ($D\ge2$) by an external potential $V(r) \propto r^m$.  
As the high-density limit sheds light on intermediate densities, the large-dimension limit provides useful insights into the $D=3$ case. \cite{Witten80,Yaffe83}
For the helium-like ions ($m=-1$), the spherium atoms \cite{Ezra82, Ezra83, Ojha87, Hinde90, Warner85, Seidl07b, TEOAS1, Quasi09, Loos10} ($m=0$), and the Hooke's law atoms \cite{Kestner62, Kais89, Taut93, Cioslowski00} ($m=2$), we found that, in the large-$D$ limit,
\begin{equation} \label{conjecture}
	\Ec^{(0)} \sim -\delta^2/8 - C \delta^3,
\end{equation}
where $\delta=1/(D-1)$ and the coefficient $C \approx 1/6$ varies slowly with $m$.  On this basis, we conjectured that Eq. \eqref{conjecture} is true for any spherically-symmetric confining external potential.

At the end of our previous work, \cite{EcLimit09} we observed that it would be highly desirable to consider $D$-ballium, the system in which the two electrons are trapped in a $D$-dimensional ball of radius $R$.  This model is a severe test of our conjecture because it corresponds to $m = \infty$.

The present study focuses mainly on the high-density regime ($R>0$ but small) and the corresponding limiting case ($R = 0$).  We report accurate results for the restricted Hartree-Fock (HF) and exact energies (Sec. \ref{sec:HF} and \ref{sec:exact}, respectively).  For the limiting case (Sec. \ref{sec:HDL}), perturbation theory is used to expand both the HF and exact energies and this allows us to determine the limiting correlation energy in $D$-ballium.  We use atomic units throughout.

The Hamiltonian of $D$-ballium is
\begin{equation} \label{H}
	\hat{H} = - \frac{\nabla_1^2}{2} - \frac{\nabla_2^2}{2} + V(r_1) + V(r_2) + \frac{1}{r_{12}},
\end{equation}
where $r_{12} = \left| \bm{r}_1 - \bm{r}_2 \right|$ is the interelectronic distance, and the external potential is defined by
\begin{equation} \label{V}
	V(r) =	\begin{cases}
						0,		&	\text{ if } r < R,	\\
					\infty,		&	\text{otherwise}.
			\end{cases}
\end{equation}
Any physically acceptable eigenfunction of \eqref{H} must satisfy the Dirichlet boundary condition 
\begin{equation} \label{Dirichlet}
	\Psi(r_1=R) = \Psi(r_2=R) = 0.
\end{equation}

\section{\label{sec:HF} Restricted Hartree-Fock approximation}

The spin-restricted HF solution, \cite{Helgaker} which is the only HF solution in the high-density regime, is given by
\begin{equation} \label{Psi-HF}
	\Psi_{\rm HF}(r_1,r_2) = \phi(r_1) \phi(r_2).
\end{equation}
If we introduce the scaled coordinate $t = r/R$, the HF orbital $\phi(t)$ is an eigenfunction of the Fock operator
\begin{equation} \label{F}
        \Hat{F} = - \frac{1}{2R^2} \nabla_t^2 + \frac{1}{R} J_\phi(t).
\end{equation}
For $S$ states in a $D$-dimensional space, \cite{Herschbach86,EcLimit09} we have
\begin{gather}
	\nabla_t^2 = \frac{d^2}{dt^2} + \frac{D-1}{t} \frac{d}{dt},																\label{nabla}	\\
        J_\phi(t) = \int_0^1 \frac{\phi(x)^2}{\max(t,x)} F\left[ \frac{3-D}{2},\frac{1}{2},\frac{D}{2},\alpha^2 \right] x^{D-1} dx,	\label{J}
\end{gather}
where $\alpha = \frac{\min(t,x)}{\max(t,x)}$ and $F$ is the hypergeometric function.

\begin{figure}
	\begin{center}
	\includegraphics[width=0.48\textwidth]{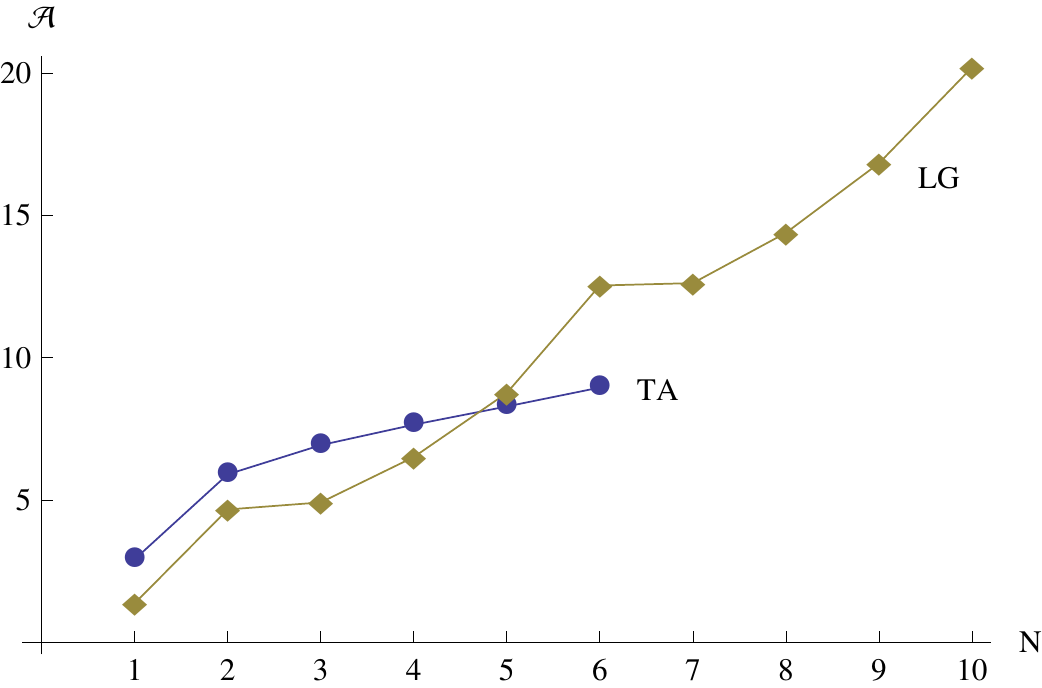}
	\caption{\label{fig:HF} Accuracy $\mathcal{A}$ of the HF energy of 3-ballium ($R=1$) with respect to basis set size $N$.  TA results taken from Ref. \onlinecite{Thompson05} and LG from the present study.}
	\end{center}
\end{figure}

Unlike Thompson and Alavi, \cite{Thompson05} who expanded the HF orbital in a basis of spherical Bessel functions, \cite{ASbook} we chose to explore an even-degree polynomial basis, writing
\begin{equation} \label{phi-HF}
	\phi(t) = \left(1-t^2\right) \sum_{k=0}^{N-1} c_k t^{2k}.
\end{equation}
Any such orbital is smooth at the center of the ball, {\em i.e.}
\begin{equation}
	\phi^{\prime}(0) = 0,
\end{equation}
and is cusped and vanishes at the boundary, {\em i.e.} 
\begin{align}
	\phi^{\prime}(1) & < 0,	&	\phi(1) & = 0.
\end{align}
It can be shown that the resulting HF energy is
\begin{equation} \label{E-HF}
	E_{\rm HF} = \frac{1}{R^2} \frac{T}{S} + \frac{1}{R} \frac{U}{S^2},
\end{equation}
with
\begin{gather}
	S = \sum_{ij} \frac{c_i c_j}{(i+j+\frac{D}{2})_3},	\\
	T = \sum_{ij} c_i c_j \left[ \frac{D}{(i+j+\frac{D}{2})_2} + \frac{4ij}{(i+j+\frac{D}{2}-1)_3} \right],	\\
\begin{split}
	U & = \sum_{ijkl} c_i c_j c_k c_l 
	\left[  \beta_{i+j+k+l-\frac{1}{2}} \gamma_{k+l} \right. \\
	& - \left. 2 \beta_{i+j+k+l+\frac{1}{2}} \gamma_{k+l+1}
	+ \beta_{i+j+k+l+\frac{3}{2}} \gamma_{k+l+2} \right],
\end{split}
\end{gather}
where 
\begin{equation}
	(a)_b = \frac{\Gamma(a+b)}{\Gamma(a)}
\end{equation}
is the Pochhammer symbol and $\Gamma$ is the Gamma function. \cite{ASbook}
The coefficients $\beta_n$ and $\gamma_n$ are given by
\begin{gather}
	\beta_n = \frac{1}{(n+D)_3},		\\
	\gamma_n = \frac{{}_3F_2\left(\frac{3-D}{2},\frac{1}{2},n+\frac{D}{2};\frac{D}{2},n+1+\frac{D}{2};1\right)}{n+\frac{D}{2}},
\end{gather}
where ${}_pF_q$ is the generalized hypergeometric function. \cite{ASbook} For $D=3$, the hypergeometric functions reduce to unity.

The energy \eqref{E-HF} can be minimized with respect to the coefficients $c_k$ using a numerical solver, \cite{Mathematica7} thus avoiding the self-consistent field procedure usually needed for this kind of calculation.\cite{Ragot08}

Henceforth, we define the accuracy of an energy $E$ as 
\begin{equation} \label{Acc}
	\mathcal{A} = - \log_{10} \left[(E-E^*)/E^*\right]
\end{equation}
where $E^*$ is our best estimate of the exact energy. In loose terms, $\mathcal{A}$ is the number of correct decimal digits.

Figure \ref{fig:HF} shows how the accuracy of the HF energy of 3-ballium ($R=1$) improves as $N$ increases.  For very small $N$, the spherical Bessel basis \cite{Thompson05} (TA) is more accurate than the polynomial (LG) basis.  However, although both the TA and LG bases seem to exhibit exponential convergence as $N$ increases, the TA energy improves by roughly one order of magnitude and the LG energy by roughly two orders of magnitude as each basis function is added.  As a result, one obtains the HF energy to 20 digits using \eqref{phi-HF} with $N=10$.  However, the origin of the superiority of the polynomial basis is not clear.  We find that the resulting expansion coefficients $c_k$ decay roughly exponentially and the convergence behavior for other $D$ is similar.

Numerical results for 3-ballium ($R=1$) are shown in Table \ref{tab:comparison}.  For $N=7$ basis functions, the Bessel and polynomial bases yield HF energies that lie 13 n$E_{\rm h}$ and 2.8 p$E_{\rm h}$ above the HF limit, respectively.  Analogous behavior is observed for the larger values of $R$, including those that lie in the low-density regime where a lower-energy UHF solution exists ($R \gtrsim 6$ for $D = 3$). \cite{Thompson05}

\begin{table*}
	\caption{\label{tab:comparison} Hartree-Fock and exact energies of 3-ballium for $R$ = 1, 5 and 20.}
	\begin{ruledtabular}
	\begin{tabular}{lclll}
			&	Basis set size	&								\mc{3}{c}{Hartree-Fock energy}										\\
																	\cline{3-5}
			&					&	\mc{1}{c}{$R=1$}				&	\mc{1}{c}{$R=5$}			&	\mc{1}{c}{$R=20$}			\\
		Thompson and Alavi \footnotemark[1]
			&		7			&	11.641 747 645				&	0.739 761 807				&	0.105 378 511				\\
		Present work using \eqref{phi-HF}
			&		7			&	11.641 747 631 859			&	0.739 761 794 626			&	0.105 378 488 0			\\
			&		10			&	11.641 747 631 855 851 828	&	0.739 761 794 625 138	&	0.105 378 488 024			\\ \\
			&					&								\mc{3}{c}{Exact energy}											\\
																	\cline{3-5}
			&					&	\mc{1}{c}{$R=1$}				&	\mc{1}{c}{$R=5$}			&	\mc{1}{c}{$R=20$}			\\
		Thompson and Alavi \footnotemark[2]
			&		210			&	11.591 380 285				&	0.701 706 934				&	0.086 577 117				\\
			&		Extrap.		&	11.590 81(4)					&	0.701 606 (2)				&	0.086 577 0(0)				\\
		Jung and Alvarellos \footnotemark[3]
			&		6296		&	11.590 906 					&	---							&	---							\\
		Present work using \eqref{Psi-xyz}
			&		196			&	11.590 838 69					&	0.701 613 820 				&	0.086 576 568 4			\\
			&		726			&	11.590 838 689 02				&	0.701 613 820 002			&	0.086 576 568 358 529	\\
	\end{tabular}
	\end{ruledtabular}
\footnotetext[1]{Reference \onlinecite{Thompson05}}
\footnotetext[2]{References \onlinecite{Thompson02} and \onlinecite{Thompson03}}
\footnotetext[3]{Reference \onlinecite{Jung03}}
\end{table*}

\section{\label{sec:exact} Explicitly correlated calculations}

We write the correlated wave function of $D$-ballium as
\begin{equation} \label{Psi-xyz}
	\Psi = \sum_{n=0}^\omega \sum_{l=0}^n \sum_{m=0}^\omega c_{nlm} \left(1+\Hat{P}_{12}\right) \psi_{nlm},
\end{equation}
where $\Hat{P}_{12}$ is the permutation operator between electron 1 and 2, which ensures the correct symmetry for the $^1S$ ground state, and the basis functions are
\begin{equation} \label{psi-nlm}
 	\psi_{nlm} = (1-x^2) (1-y^2) x^{2n} y^{2l} z^m,
\end{equation}
the scaled coordinates are
\begin{align}
	x	& = \frac{r_1}{R},	&	y	& = \frac{r_2}{R},	&	z	& = \frac{r_{12}}{R},
\end{align}
and $n$, $l$ and $m$ are non-negative integers.  Such functions ensure that $\Psi$ is smooth at $r_1 = 0$ and $r_2 = 0$, {\em i.e.}
\begin{equation}
	\left. \frac{\partial \Psi}{\partial r_1} \right|_{r_1=0} = \left. \frac{\partial \Psi}{\partial r_2} \right|_{r_2=0} = 0,
\end{equation}
and that $\Psi$ is cusped at the boundary and satisfies \eqref{Dirichlet}.  The total number of basis functions in \eqref{psi-nlm} is
\begin{equation}
	N = \frac{(\omega+1)^2(\omega+2)}{2}.
\end{equation}

The ground-state energy is the lowest eigenvalue of
\begin{equation}
	\mathbf{S}^{-1/2} (\mathbf{T} + \mathbf{U}) \mathbf{S}^{-1/2},
\end{equation}
where $\mathbf{S}$, $\mathbf{T}$ and $\mathbf{U}$ are the overlap, kinetic and repulsion matrices, respectively. \cite{Hylleraas64}

Although our $(x,y,z)$ coordinates are equivalent to the $(s,t,u)$ coordinates of Hylleraas, \cite{Hylleraas29, Hylleraas64} ours lead to simpler closed-form expressions for the required integrals.  All the required matrix elements can be found in closed form using the general formula in Appendix \ref{app:int}.

\begin{figure}
	\begin{center}
       \includegraphics[width=0.48\textwidth]{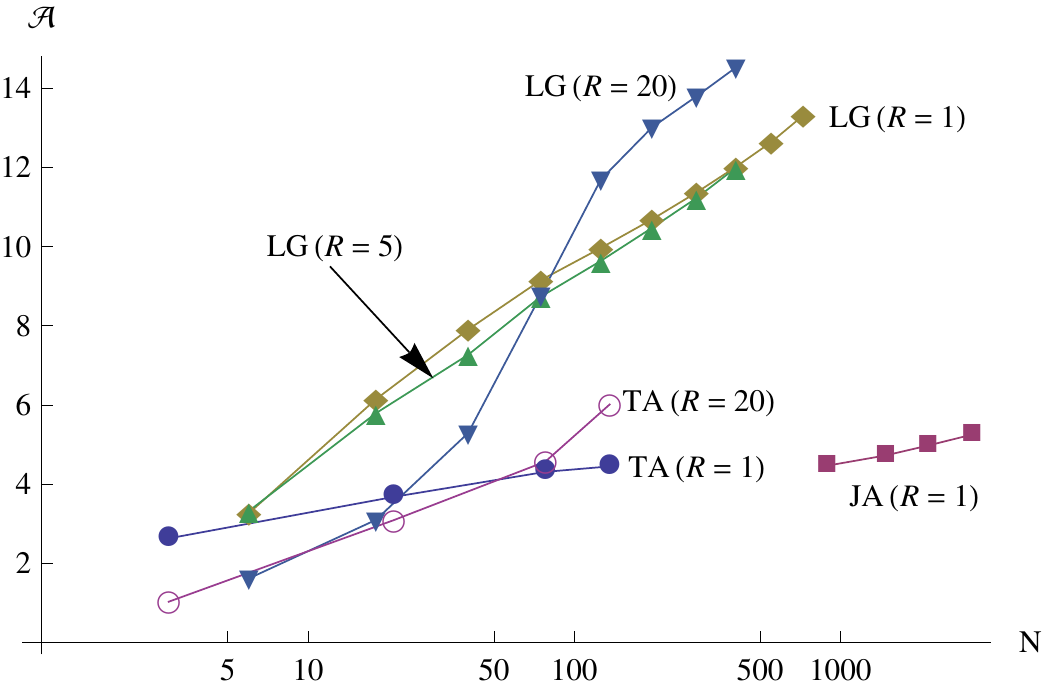}
	\caption{\label{fig:exact} 
Accuracy $\mathcal{A}$ of the exact energy of 3-ballium with respect to the basis set size $N$ for various $R$.  TA results taken from Ref. \onlinecite{Thompson02}, JA from Ref. \onlinecite{Jung03} and LG from the present study.}
	\end{center}
\end{figure}

Figure \ref{fig:exact} shows how the accuracy of the exact energy of 3-ballium improves as the number $N$ of terms in the expansion increases and Table \ref{tab:comparison} reports numerical values of the exact energy for various $R$ = 1, 5 and 20.  Our explicitly correlated results are compared with the CI energies of Thompson and Alavi \cite{Thompson02, Thompson03} and of Jung and Alvarellos. \cite{Jung03}  Convergence for other values of $D$ is similar.

Explicitly correlated calculations converge much faster than CI calculations because the former include terms \eqref{psi-nlm} with $m=1$ satisfying the Kato cusp condition. \cite{Kato57, Kutzelnigg85, Kutzelnigg91}  For example, for the unit ball, Thompson and Alavi \cite{Thompson02, Thompson03} obtained $E$ = 11.591 380 285 using 210 basis functions, Jung and Alvarellos \cite{Jung03} subsequently found $E$ = 11.590 906 using 6296 functions, but we obtain $E$ = 11.590 838 689 using only 196 functions ($\omega = 6$).  Our energy is consistent with the extrapolated \cite{Halkier98} estimate $E$ = 11.590 81(4) of Thompson and Alavi. \cite{Thompson02, Thompson03}  Likewise, using 6296 basis functions, Jung and Alvarellos \cite{Jung03} found $E$ = 22.033 71 for the first excited $S$ state and, using 196 explicitly correlated functions, we obtain $E$ = 22.033 562 4 $E_{\rm h}$.

Figure \ref{fig:exact} reveals that, for $R=1$ and $R=5$, the rate of convergence of the Hylleraas basis set is very similar.  When $R$ exceeds the Wigner-Seitz critical value ($r_s  \approx 6$ for $D=3$), a Wigner molecule is formed, characterized by a minimum of the electron density at the center of the box. \cite{Thompson02, Thompson03}  Although the CI calculations of Thompson and Alavi show only small variations of the rate of convergence, the Wigner molecule formation dramatically modifies the energy convergence of our explicitly correlated calculations.  For $R=20$ and short expansions ($N < 50$), the CI and Hylleraas calculations lead to similar results but, for larger basis sets, the Hylleraas scheme is superior and the accuracy rapidly improves.

Correlation energies $\Ec$ for $D = 2, 3,\ldots,7$ and $R$ = 1, 5 and 20 are given in Table \ref{tab:E2}.  For fixed $D$ and increasing $R$, $\Ec$ decreases.  For fixed $R$ and increasing $D$, although both the exact and HF energies increase, $\Ec$ decreases. \cite{Loeser86a,Loeser86b,Loeser87b}

\begin{table*}
\caption{\label{tab:E2} Exact, HF and correlation energies for various finite $R$.  Zeroth-, first-, second-order energies and limiting correlation energies.}
\begin{ruledtabular}
\begin{tabular}{lcccccc}
	$D$				&		2		&		3		&		4		&		5		&		6		&		7			\\
	\hline
						&		\mc{6}{c}{Exact, HF and correlation energies of ballium for $R=1$}							\\
	$E$				&	8.104931	&	11.590839	& 	16.151742	&	21.519813	&	27.612654	&	34.391191		\\
	$E_{\rm HF}$		&	8.326496	&	11.641748	& 	16.172654	&	21.530902	&	27.619443	&	34.395746		\\
	$\Ec$				&	-0.221565	&	-0.050909	&	-0.020912	&	-0.011089	&	-0.006789	&	-0.004555		\\	\\
						&		\mc{6}{c}{Exact, HF and correlation energies of ballium for $R=5$}							\\
	$E$				&	0.586796	&	0.701614	&	0.863437	&	1.063334	&	1.296170	&	1.559045		\\
	$E_{\rm HF}$		&	0.711077	&	0.739762	&	0.880997	&	1.073192	&	1.302407	&	1.563317		\\
	$\Ec$				&	-0.124281	&	-0.038148	&	-0.017560	&	-0.009858	&	-0.006237	&	-0.004272		\\	\\
						&		\mc{6}{c}{Exact, HF and correlation energies of ballium for $R=20$}						\\
	$E$				&	0.078628	&	0.086577	&	0.096381	&	0.108129	&	0.121815	&	0.137388		\\
	$E_{\rm HF}$		&	0.123044	&	0.105378	&	0.107060	&	0.114985	&	0.126552	&	0.140835		\\
	$\Ec$				&	-0.044416	&	-0.018801	&	-0.010679	&	-0.006856	&	-0.004737	&	-0.003447		\\	\\
						&		\mc{6}{c}{Zeroth- and first-order energies of ballium, from Eq. \eqref{E-0-1}}				\\
	$E^{(0)}$			&	5.783186	&	9.869604	&	14.681971	&	20.190729	&	26.374616	&	33.217462		\\
	$E^{(1)}$			&	2.596157	&	1.786073	&	1.496754	&	1.343463	&	1.246845	&	1.179626		\\	\\
						&		\mc{6}{c}{Second-order energies of ballium, from Eqs. \eqref{E2-HF} and \eqref{E2}}		\\
	$E^{(2)}$			&	-0.324120	&	-0.069618	&	-0.028107	&	-0.014770	&	-0.008977	&	-0.005983 		\\
	$E_{\rm HF}^{(2)}$	&	-0.057959	&	-0.014442	&	-0.006194	&	-0.003333	&	-0.002037	&	-0.001352		\\	\\
						&		\mc{6}{c}{Limiting correlation energies $\Ec^{(0)}$, from Ref. \onlinecite{EcLimit09} and Eq. \eqref{Ec-0}}	\\
	Helium				&	-0.220133	&	-0.046663	&	-0.018933	&	-0.010057	&	-0.006188	&	-0.004176		\\
	Spherium			&	-0.227411	&	-0.047637	&	-0.019181	&	-0.010139	&	-0.006220	&	-0.004189		\\
	Hookium			&	-0.239641	&	-0.049703	&	-0.019860	&	-0.010439	&	-0.006376	&	-0.004280		\\
	Ballium				&	-0.266161	&	-0.055176	&	-0.021913	&	-0.011437	&	-0.006940	&	-0.004631		\\
\end{tabular}
\end{ruledtabular}
\end{table*}

\section{\label{sec:HDL} Limiting correlation energy}

Following Hylleraas perturbation theory, \cite{Hylleraas30} we expand both the exact and HF energies as series in $R$, yielding
\begin{gather}
	E = \frac{E^{(0)}}{R^2} + \frac{E^{(1)}}{R} + E^{(2)} + O(R),		\\
	E_{\rm HF} = \frac{E^{(0)}}{R^2} + \frac{E^{(1)}}{R} + E_{\rm HF}^{(2)} + O(R).
\end{gather}
The limiting correlation energy is then given by
\begin{equation} \label{Ec-0}
	\Ec^{(0)} = \lim_{R \to 0} \Ec = E^{(2)} - E_{\rm HF}^{(2)}.
\end{equation}
The one-electron Hamiltonian for $D$-ballium is
\begin{equation} \label{H0}
	\Hat{H}_0 = - \frac{1}{2} \left[ \frac{d^2}{dr^2} + \frac{D-1}{r} \frac{d}{dr} \right] + V(r),
\end{equation}
and the associated zeroth-order wave function is
\begin{equation} \label{Psi0}
	\Psi_0(r_1,r_2) = \psi_0(r_1) \psi_0(r_2),
\end{equation}
where
\begin{equation} \label{psi0}
	\psi_0(r) = \frac{\sqrt{2}}{J_{D/2}(\kappa)} \frac{J_{D/2-1}(\kappa r)}{r^{D/2-1}},
\end{equation}
In \eqref{psi0}, $\kappa = j_{D/2-1,1}$ and $j_{\mu,k}$ is the $k$-th zero of the Bessel function of the first kind $J_{D/2-1}$. \cite{ASbook}  The $E^{(0)}$ and $E^{(1)}$ values are easily obtained from the relations
\begin{align} \label{E-0-1}
	E^{(0)} & = \kappa^2,		&	E^{(1)} & = \left< \Psi_0 \left| r_{12}^{-1} \right| \Psi_0 \right>,
\end{align}
and are reported in Table \ref{tab:E2}.  For odd D, $E^{(1)}$ can be found in closed form.  For example, for $D=3$,
\begin{equation}
	E^{(1)} = 2 \left[1 - \frac{\Si(2\pi)}{2\pi} + \frac{\Si(4\pi)}{4\pi}\right],
\end{equation}
where $\Si$ is the sine integral function. \cite{ASbook}

\subsection{\label{subsec:HDL-HF} Hartree-Fock energy}

Values of $E_{\rm HF}^{(2)}$ have been determined using the generalization of the Byers-Brown--Hirschfelder equations \cite{ByersBrown63}
\begin{gather}
	E_{\rm HF}^{(2)} = - \int_0^1 \frac{W(r)^2}{r^{D-1} \,\psi_0(r)^2}\,dr,	\label{E2-HF}	\\
	W(r) = 2 \int_0^r [J_{\psi_0}(x) - E^{(1)}] \,\psi_0(x)^2 \, x^{D-1} \, dx,
\end{gather}
where $J_{\psi_0}(x)$ is given by \eqref{J}. 

\subsection{\label{subsec:HDL-exact} Exact energy}

The second-order energy $E^{(2)}$, which minimizes the Hylleraas functional, \cite{Hylleraas30} is given by
\begin{equation} \label{E2}
	E^{(2)} = - \mathbf{b^{\rm T}A^{-1}b},
\end{equation}
where
\begin{gather}
	\mathbf{A} = \mathbf{T} - E^{(0)} \mathbf{S},	\\
	\mathbf{b} = \mathbf{C}^{\rm T} \left[ E^{(1)} \mathbf{S} - \mathbf{U} \right].
\end{gather}
The matrices $\mathbf{S}$, $\mathbf{T}$ and $\mathbf{U}$ have been defined in Sec. \ref{sec:exact}.  The vector $\mathbf{C}$ contains the coefficients of the zeroth-order wave function \eqref{Psi0} expanded in the basis set \eqref{psi-nlm}.  The basis set has been enlarged by progressively increasing the maximum value of $\omega$.

\subsection{\label{subsec:HDL-Ec} Correlation energy}

The exact and HF second-order energies, as well as the limiting correlation energy $\Ec^{(0)}$, of $D$-ballium are reported in Table \ref{tab:E2}.  The latter is compared with previously reported results \cite{EcLimit09} for related two-electron systems (helium, spherium and hookium).

The first observation is the tendency of the limiting correlation energies to decrease with increasing dimensionality. \cite{Herrick75}  As $D$ increases, all of the energies decrease dramatically and the correlation energies fall by almost two orders of magnitude between $D = 2$ and $D = 7$.  This point has been already discussed and explained in previous works. \cite{Herrick75,EcLimit09}

We have used the method developed by Herschbach and collaborators \cite{Loeser85, Doren85, Herschbach86, Loeser87a, Goodson87} to find that the large-$D$ expansion of $\Ec^{(0)}$ in $D$-ballium is
\begin{equation} \label{Ec-ba}
	\Ec^{(0)} \sim - \frac{1}{8} \delta^2 - \frac{53}{128}\delta^3 + \ldots,
\end{equation}
where $\delta = 1/(D-1)$. \cite{Yaffe83}  This supports our recent conjecture \eqref{conjecture} that the leading term $-\delta^2/8$ is universal and independent of the radial external potential $V(r)$.  
We note that the coefficient of $\delta^3$ is larger than those in the other two-electron systems and this explains why the limiting correlation energy of $D$-ballium is always larger than those in helium, spherium and hookium.

\section{\label{sec:ccl} Conclusion}

In this article, we have reported accurate results for the exact and HF ground-state energies of two electrons of opposite spin confined within a ball of radius $R$ in a $D$-dimensional space.  We call this system $D$-ballium.

Our results, focussing mainly on the high-density regime (small-$R$) and the limit where $R=0$, extend and complete earlier studies on 3-ballium. \cite{Thompson02, Jung03, Thompson05}  The exact results have been obtained using a Hylleraas-type expansion, while the HF limit has been reached using a simple polynomial basis set.

We have also shown that, in the large-dimension limit, the limiting correlation energy behaves as $-\delta^2/8 - C \delta^3$, in agreement with our recent conjecture, \cite{EcLimit09} and is larger than the limiting correlation energy in other two-electron systems.  
A rigorous proof of the conjecture will be reported elsewhere, \cite{proof10} where we will show that this result is related to the Kato cusp factor.  Using continuity arguments, it seems clear that the conjecture does not apply to non-spherical external potentials.

\begin{acknowledgments}
PMWG thanks the NCI National Facility for a generous grant of supercomputer time and the Australian Research Council (Grant DP0984806) for funding.
\end{acknowledgments}

\appendix

\section{\label{app:int} Hylleraas-type integrals}

The integrals needed to compute the different matrix elements are of the form
\begin{equation} \label{Int}
	\mathcal{I}_{nlm} = \int x^n y^l z^m d\tau,
\end{equation}
with the volume element
\begin{gather}	
	d\tau = x\,y\,z\,\mathcal{J}^\frac{D-3}{2}\,dx\,dy\,dz 	\label{dV},	\\
	\mathcal{J} = (x+y+z)(x-y+z)(x+y-z)(x-y-z),
\end{gather}
and domain of integration
\begin{equation} \label{intdV}
	\int d\tau = \int_0^1 dx \int_0^1 dy \int_{|x-y|}^{x+y} dz.
\end{equation}
One eventually finds
\begin{equation}
	\mathcal{I}_{nlm} = \sqrt{\pi}\frac{\Gamma\left(\frac{D-1}{2}\right)}{\Gamma\left(\frac{D}{2}\right)} 
	\frac{R^{n+l+m+2D}}{n+l+m+2D} \left(I_n^m+I_l^m\right),
\end{equation}
and
\begin{equation}
	I_a^b = \frac{{}_3F_2\left(\frac{a+D}{2},-\frac{b}{2},-\frac{b+D-2}{2}; \frac{a+D+2}{2},\frac{D}{2};1\right)}{a+D}.
\end{equation}

%

\end{document}